\documentclass[12pt]{iopart}
\usepackage{iopams}
\usepackage{epsfig}

\begin{document}

\title[Percolation in deposits for competitive models in (1+1)-dimensions]
{Percolation in deposits for competitive models in
(1+1)-dimensions}
\author{
N I Lebovka\dag\footnote[3]{To whom correspondence should be
addressed.} \footnote[4]{E-mail: lebovka@roller.ukma.kiev.ua}, S S
Manna\ddag , S Tarafdar\P  and N V Vygornitskii\dag
}

\address{\dag\ Institute of Biocolloid Chemistry NASU, 2, bulv.
Vernadskogo, Kyiv, 03142, Ukraine}

\address{\ddag\ Satyendra Nath Bose National
Centre for Basic Sciences Block-JD, Sector-III, Salt Lake,
Kolkata-700098, India}

\address{\P\ Condensed Matter Physics Research
Centre, Physics Department, Jadavpur University, Kolkata 700032,
India}

\begin{abstract}
The percolation behaviour during the deposit formation, when the
spanning cluster was formed in the substrate plane, was studied.
Two competitive or mixed models of surface layer formation were
considered in (1+1)-dimensional geometry. These models are based
on the combination of ballistic deposition (BD) and random
deposition (RD) models or BD and Family deposition (FD) models.
Numerically we find, that for pure RD, FD or BD models  the mean
height of the percolation deposit $\bar h$ grows with the
substrate length $L$ according to the generalized logarithmic law
$\bar h\propto (\ln (L))^\gamma$, where $\gamma=1.0$ (RD),
$\gamma=0.88\pm 0.020$ (FD) and $\gamma=1.52\pm 0.020$ (BD). For
BD model, the scaling law between deposit density $p$ and its mean
height $\bar h$ at the point of percolation of type $p-p_\infty
\propto \bar h^{-1/\nu_h}$ are observed, where $\nu_h
=1.74\pm0.02$ is a scaling coefficient. For competitive models the
crossover, 
corresponding to the RD or FD -like behaviour  at small $L$ and
the BD-like behaviour at large $L$ are observed.
\end{abstract}

 \pacs{64.60.Ak, 68.00.00, 72.60.+g, 81.15.Aa,
89.75.Da}


\submitted

\maketitle

\section{Introduction}
\label{Introduction}
Formation of thin films in the process of deposition and
aggregation of particles on a substrate has received a
considerable theoretical and experimental attention in recent
years \cite{Family91,Barabasi95}. This problem is very important
also from the practical point of view for production of thin-film
devices and conducting composite films with certain specified
electrical \cite {Electrical1,Electrical2}, magnetic, transport
\cite{Magnetic} and colloidal properties \cite
{Colloid1,Colloid2}. There also exist large interest in studying
morphology of deposits \cite{Morphology1,Morphology2,Morphology3},
their fractal \cite {Fractal1,Fractal2} and percolating properties
\cite {Percolation1,Percolation2,Percolation3,Percolation4}.

Among the most popular models for simulation of deposits formation
there are models of random deposition (RD), random deposition with
surface relaxation or Family deposition (FD) and ballistic
deposition (BD) and different their variants
\cite{Family91,Barabasi95}. In these models, the particles are
rigid and cannot overlap and these models describe growth
processes far from equilibrium. In RD model the particles deposit
without sticking and it means the presence of the short-range
repulsion. In FD model the particles can relax to a lower nearest
neighbor position. In BD model the particles stick at a point of
the first contact and it means the presence of the short-range
attraction. Recently there were proposed a number of mixed or
competitive models, which are based on consideration of deposition
from different kinds of particles \cite{ModelMixed1,ModelMixed2}.

The percolation phenomena in growing simulated films were
previously analyzed for different models of deposits on
two-dimensional substrate
\cite{ModelPerco1,ModelPerco2,ModelPerco3,ModelPerco4,Lebovka02}.
The present paper addresses the percolation behaviour for
different competitive (1+1)-dimensional lattice models of deposit
formation on a line substrate.

The paper is organized as follow. The model is described in
section \ref{model}. In section \ref{results}, the scaling
behaviour of deposit height at percolating point and its density
are discussed. Concluding remarks are presented in section
\ref{conclusions}

\section{Model}
\label{model}

In our  (1+1)-dimensional competitive models there exist two kinds
of particles, following BD rules and RD or FD rules. We call these
models as BD$_{1-s}$RD$_{s}$ or BD$_{1-s}$FD$_{s}$, respectively,
where $s$ is a fraction of RD or FD particles. Each particle falls
along vertical direction, until it reaches the interface.
Particles get deposited one after another and fixed in the sites
of square lattice according to the deposition rules.

We stop the growth process when the spanning cluster forms for the
first time in the substrate plane and this point is easily checked
by a Hoshen-Kopelman algorithm \cite{HK}. The percolation in
deposits differs from usual random percolation \cite{Stauffer92}
and has a correlated character, because the sites of lattice get
filled dynamically during the growth of deposit.

The mean height of deposit at the percolation point is calculated
as $ \bar h=\Sigma_{i}h_{i}/L$, where $L$ is the substrate length.
The time $t$ is counted as the number of deposited particles
$N=L$. The deposit density $p$ is calculated as the ratio of the
number of particles and the deposit volume $p=N/(L \bar h)=t/\bar
h$.

The substrate size $L$ was varied from $2^2$ to $2^{17}$ and the
periodical boundary conditions were used in the deposition rules
along the substrate direction. Results were averaged over 100-5000
different runs, depending on the size of the lattice and required
precision.

\section{Results and discussion}
\label{results}

\begin{figure}[top]
\centerline{\epsfxsize=5in \epsfbox{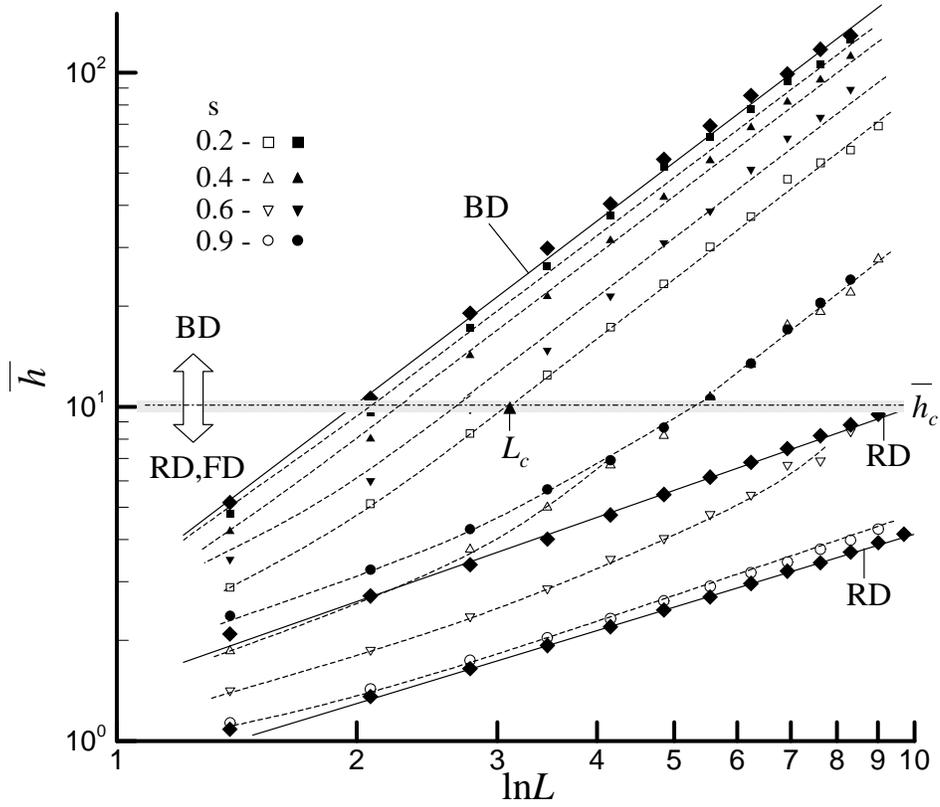}}
\caption{Mean height of deposit at the percolation point $\bar h$
versus substrate length $L$ for BD$_{1-s}$ FD $_{s}$ (open
symbols) and BD $_{1-s}$ RD $_{s}$ (filled symbols) models at
different $s$. Here the crossover line $\bar h \approx 10$ between
regimes with different $\bar h$ versus $L$ behaviour is shown. The
data error is of the order of data symbol size. The dashed lines
serve as a guide to the eye. Both of $\bar h$ and $L$ are in
lattice units.} \label{f01}
\end{figure}

In \fref{f01} we show the numerically determined mean height of
the deposit at the percolation point $\bar h$ versus the substrate
length $L$ for competitive models BD$_{1-s}$RD$_{s}$ and
BD$_{1-s}$FD$_{s}$ at various values of $s$. A remarkable feature
of data presented in \fref{f01} is that for for pure RD($s=1$),
FD($s=1$) or BD($s=0$) models,  the value of $h$ grows with the
substrate length $L$ according to the generalized logarithmic law:
\begin{equation} \label{ln-gamma}
h\propto (\ln L)^\gamma,
\end{equation}
where $\gamma$ is an exponent that is  $\gamma_{RD}=1.0$,
$\gamma_{FD}=0.88\pm 0.02$ and $\gamma_{BD}=1.52\pm 0.02$ for RD,
FD and BD models, respectively.

For RD model the logarithmic law $\bar h=\ln L$ can be easily
justified. In the RD model, every column of deposit grows
independently, and in the limit of $L\gg1$ the distribution
function of heights follows the Poisson law:
\begin{equation}
P(h)=(e^{-\bar h})({\bar h}^{h})/h!.
\end{equation}

\begin{figure}[top]
\centerline{\epsfxsize=5in \epsfbox{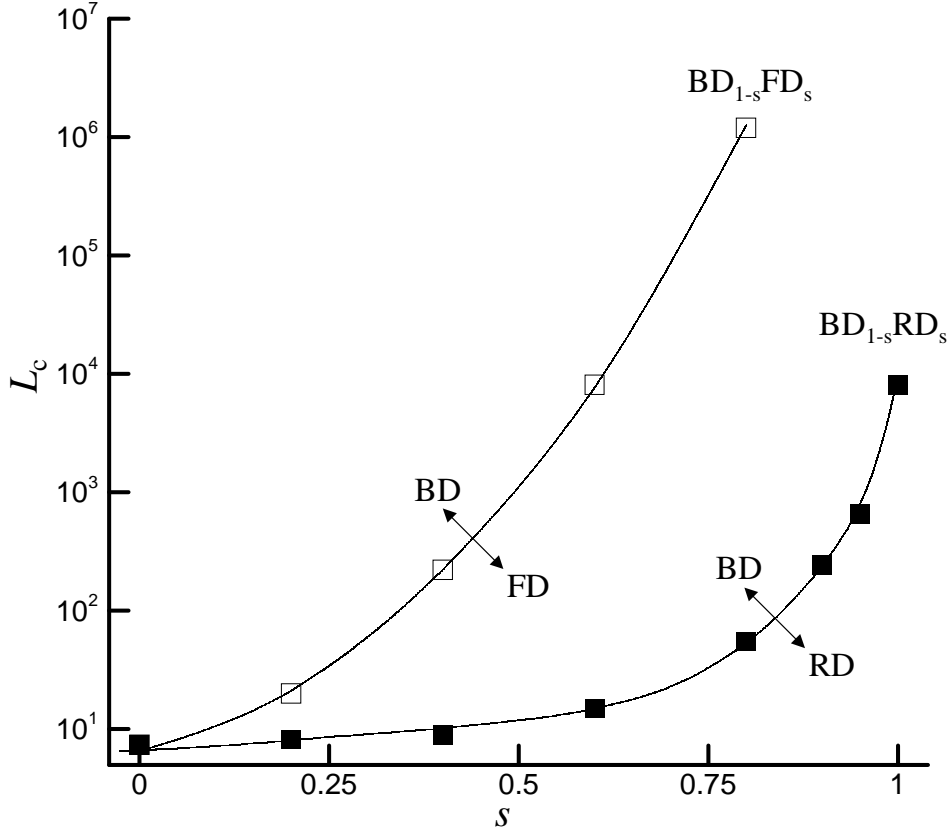}}
 \caption{Crossover
length $L_c$  versus $s$ for BD$_{1-s}$ FD $_{s}$ and BD $_{1-s}$
RD $_{s}$ models.} \label{f02}
\end{figure}

Therefore, the probability to find of an empty column ($h=0$) is
equal to $P(0)=e^{-\bar h}$. In the percolation point the last
empty column gets filled, and $P(0)=e^{-\bar h}=1/L$. So, we
obtain for RD model exactly
\begin{equation}
\bar h = \ln L,
\end{equation}
and therefore $\gamma=1$ in \eref{ln-gamma} for that model.

For pure BD and FD model, there exist correlations between the
columns and exact relation can not be obtained by this simple way.
For both RD and FD models, the deposits are non-porous and $p=1$
at any time during the deposit formation. Moreover, in the
percolation point at the same $L$, the mean height of the deposit
is less for FD model than for RD model, i.e., $\bar h_{FD}<\bar
h_{RD}$.  For BD model, the deposit is porous and, therefore,
$\bar h_{RD}<\bar h_{BD}$ at the same $L$. So, it is reasonable to
expect that $\gamma_{FD}<\gamma_{RD}<\gamma_{BD}$ and it is
completely in accordance with numerical data presented in
\fref{f01}.

For competitive BD$_{1-s}$ FD $_{s}$ and BD $_{1-s}$ RD $_{s}$
models, the data follow the generalized logarithmic law
\eref{ln-gamma} only at large values of $\bar h$ and systematic
deviations from this law are observed $\bar h\lesssim \bar h_c$,
where $\bar h_c \approx 10$. We attribute the origin of this
behaviour to the known random-like nature of the deposition
process at small $\bar h$, when there are no strong correlations
between growth columns in the deposit, and the choice of the
boundary value $h_c$ is rather approximate \cite{ModelMixed2}.
\begin{figure}[top]
\centerline{\epsfxsize=5in \epsfbox{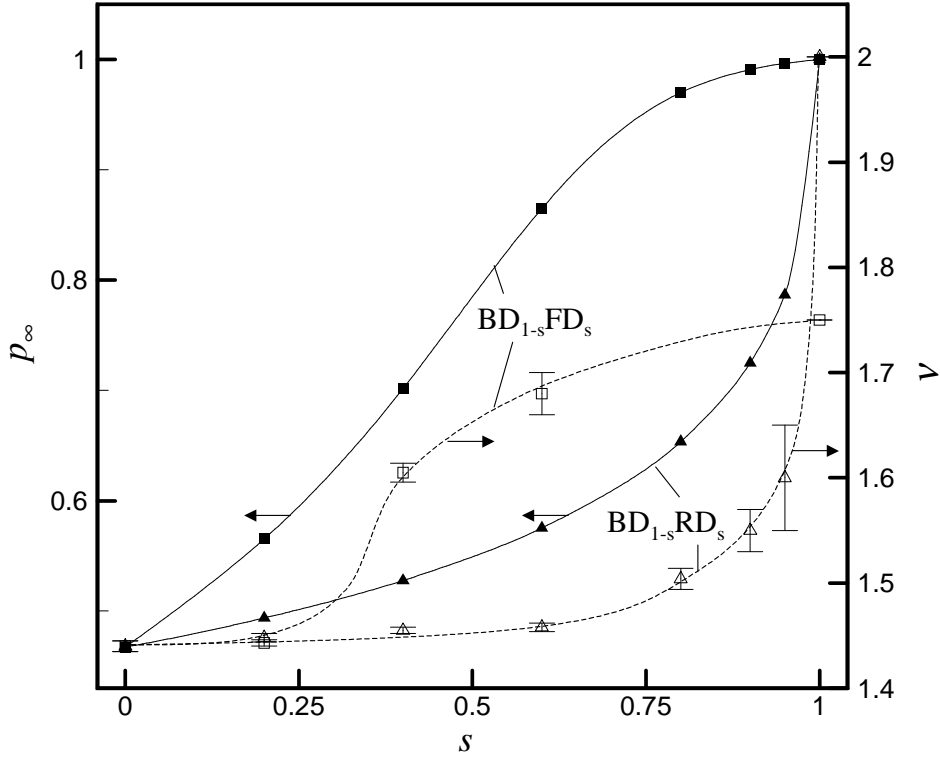}}
 \caption{Plots of
density of deposit $p_\infty$ and scaling exponent $\nu$ in
\eref{p-L} versus $s$ for BD$_{1-s}$ FD $_{s}$ and BD $_{1-s}$ RD
$_{s}$ models. In cases when it is not show directly the data
error is of order of data symbol size. The lines serve as a guide
to the eye.} \label{f03}
\end{figure}

So, for the competitive model with a given $s$, the random-like
processes can control the $\bar h$ versus $L$ behaviour at small
$L$ and there exists a strong crossover to the BD-like behaviour
at $L>L_c$, where $L_c$ is the crossover length.  We have
estimated the value of $L_c$ from the intersection point of $\bar
h(L)$ line with the horizontal line $\bar h =h_c \approx 10$. The
crossover length $L_c$ is always smaller for BD $_{1-s}$ RD $_{s}$
model than $L_c$ for BD $_{1-s}$ FD $_{s}$ model at the same $s$
(see \fref{f02}).

It is interesting to check for existence of some scaling between
the value of  the deposit density $p$ and its height $\bar h$ in
the percolation point. In order to do this analysis correctly, we
need for a limiting value of a deposit density $p_\infty$ for
infinitely large systems. We have analyzed the $p$ versus $L$
dependencies for the systems of size $L \times \bar h$, with $\bar
h=L$ and have found that all the data can be fitted with the
following scaling relation:
\begin{equation}
p-p_\infty \propto L ^{-1/\nu}, \label{p-L}
\end{equation}
where $p_\infty$ is the deposit density in the limit of
$L\rightarrow \infty$, and $\nu$ is an exponent.

For BD model $p_{\infty, BD}=0.4673$ (this result is in accordance
with \cite{Krug91}) and $\nu_{BD} = 1.44 \pm 0.01$, and $p_\infty$
and $\nu$ increase with $s$ increase for both BD$_{1-s}$ FD $_{s}$
and BD $_{1-s}$ RD $_{s}$ models (\fref{f03}).
\begin{figure}[top]
\centerline{\epsfxsize=5in \epsfbox{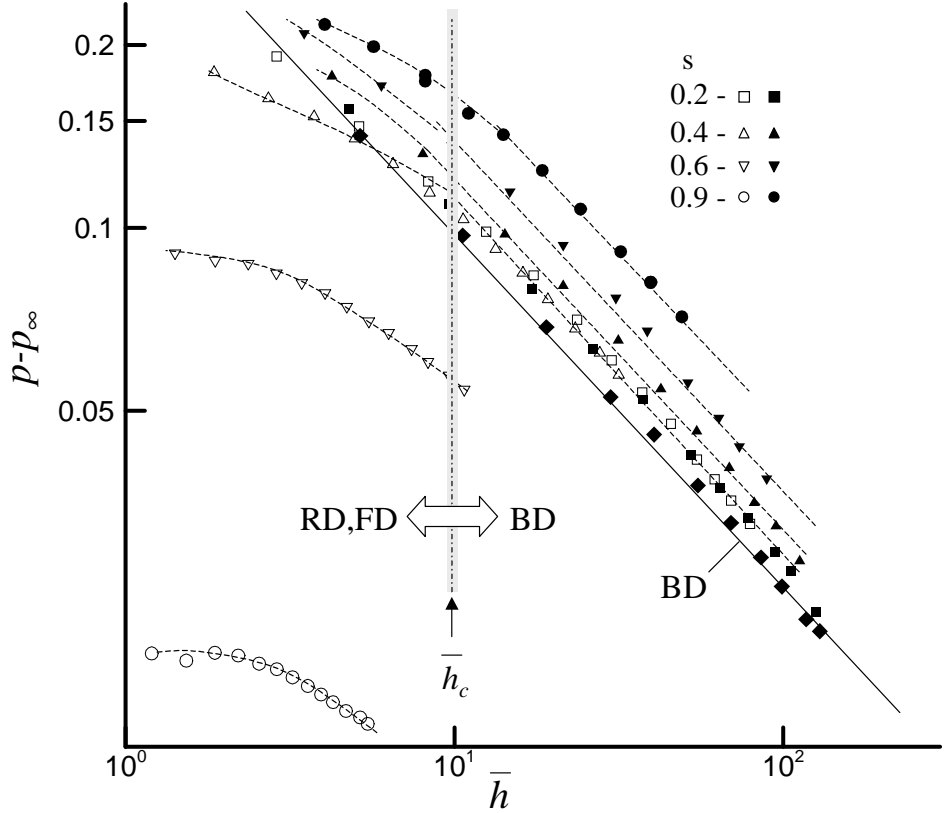}}
 \caption{Log-log
plot of $p-p_{\infty}$ versus $\bar h$ (lattice units) for
BD$_{1-s}$ FD $_{s}$ (open symbols) and BD $_{1-s}$ RD $_{s}$
(filled symbols) models at different $s$. Here the crossover line
$\bar h \approx 10$ is shown. The data error is of order of data
symbol size. The dashed lines serve as a guide to the eye. The
solid line corresponds to best fit of \eref{Scaling-ph} to the
data for BD model(filled diamonds) with scaling exponent
$\nu_{h}=1.74 \pm 0.02$.} \label{f04}
\end{figure}

We can estimate the values of $\nu_{RD}$ and $\nu_{FD}$ from the
following reasonings. The number of the deposited particles $N$ in
the system of size $L\times \bar h=L\times L$ with the rough upper
interface can be estimated as $N=L(L-a\omega)$, where $a$ is some
constant and $\omega$ is the interface width. The value of
$\omega$ scales at small times $t=\bar h$ as $\omega \propto
t^\beta$, where $\beta$ is the growth exponent which is equal to
$1/2$ for RD model and to $1/4$ for FD model \cite{Barabasi95}.
For compact deposits $p=1$, $t=\bar h$ and, so, $\omega \propto
\bar h^\beta=L^\beta$. For both RD and FD models, the height of
deposit in our systems $\bar h=L$ corresponds to the the
small-time regime, because the time of transition to the
saturation regime is $t_x= \infty $ for the RD model and of order
$t_x \approx L^2$ for the FD model. So, the density of the deposit
at $s\rightarrow 1$ can be estimated as:
\begin{equation}
p=N/L^2=L(L-a\omega)/L^2=1-aL^{-1/\nu},
\end{equation}
where $\nu=1/(1-\beta)$, and $\nu_{RD}=2$ and $\nu_{RD}=4/3$ for
RD and FD models, respectively.

\Fref{f04} shows the log-log presentation of $p-p_\infty$ versus
$\bar h$ for the BD$_{1-s}$ FD $_{s}$ and BD $_{1-s}$ RD $_{s}$
models. The evident scaling of type
\begin{equation}
p-p_{\infty}\propto \bar h^{-1/\nu_h}, \label{Scaling-ph}
\end{equation}
is observed outside the random-like deposition regime at $\bar
h\gtrsim 10$ for both BD$_{1-s}$ FD $_{s}$ and BD $_{1-s}$ RD
$_{s}$ models. At $\bar h\gtrsim 10$, the scaling exponent at
different $s$ was approximately same as for pure BD model
$\nu_h=1.74 \pm 0.02$.

\section{Conclusions}
\label{conclusions}

In summary, we have investigated the percolation in the direction
parallel to the surface for competitive (1+1) dimensional models
of deposition layer formation. The height of the percolating
deposit $\bar h$ shows the continuous growth with increase of the
substrate length $L$, and percolation is absent in the limit of
infinite systems  $L\rightarrow \infty$. This behaviour is
different from that observed for (2+1) dimensional model, where a
percolating deposit with finite height was formed
\cite{Lebovka02}. The competitive models always show a crossover
to the BD-like deposition behaviour at the limit of very large
$L$.


\section*{Acknowledgements}
NL and NV thanks the NASU for partial financial support under the
projects No. 2.16.1.4 (0102V007058) and 2.16.2.1(0102V007048).


\end{document}